\documentclass{PoS}
\usepackage{caption}
\usepackage{subcaption}
\usepackage{wrapfig}
\newcommand{\s }{$S_{125}$}

\title{Cosmic Ray Spectrum and Composition from PeV to EeV from the IceCube Neutrino Observatory}

\ShortTitle{IceCube Cosmic Ray Spectrum and Composition }

\author{

The IceCube Collaboration\footnote{For collaboration list, see PoS(ICRC2019) 1177.}\\
{\itshape \href{http://icecube.wisc.edu/collaboration/authors/icrc19_icecube}{http://icecube.wisc.edu/collaboration/authors/icrc19\_icecube}}\\
E-mail: \email{karen.andeen@marquette.edu, matthias.plum@marquette.edu}
}

\abstract{
The IceCube Neutrino Observatory at the South Pole is a multi-component detector capable of measuring the cosmic ray energy spectrum and composition from PeV to EeV, the energy region typically thought to cover the transition from galactic to extragalactic sources of cosmic rays.  The IceTop array at the surface is sensitive to the electromagnetic part of the air shower while the deep in-ice array detects the high-energy (TeV) muonic component of air showers.  IceTop's reconstructed shower size parameter, S$_{125}$, is unfolded into a high statistics all-particle energy spectrum. Furthermore, for air showers that pass through both arrays, the in-ice reconstructed muon energy loss information is combined with S$_{125}$ in a machine learning algorithm to simultaneously extract both the all-particle energy spectrum and individual spectra for elemental groups. The all-particle spectra as well as spectra for individual elemental groups are presented.\\

\vspace{4mm}
{\bfseries Corresponding authors:}
\speaker{Karen Andeen}$^{1}$, Matthias Plum$^{1}$\\
{$^{1}$ \itshape Marquette University}\\
}

\FullConference{36th International Cosmic Ray Conference -ICRC2019-\\
		July 24th - August 1st, 2019\\
		Madison, WI, U.S.A.}

\begin{document}

\section{Introduction}
In December of 2010, the IceCube Neutrino Observatory (IceCube) was completed, marking the dawn of a new era in neutrino astronomy \cite{IceCube_detector}.  IceCube is also a world-class cosmic ray observatory, sensitive to both the energy spectrum and composition of cosmic rays at energies from PeV to EeV \cite{icetop_technical}.  Although the sources, acceleration and propagation of high-energy cosmic rays are not well-understood, the PeV to EeV energy regime is particularly interesting because it may cover the transition from galactic to extragalactic cosmic ray sources (as discussed in \cite{KAMPERT2012660}, for example).

\begin{figure}[tb]
\centering
\hspace*{\fill}
\begin{subfigure}[t]{0.35\textwidth}
\includegraphics[width=\textwidth]{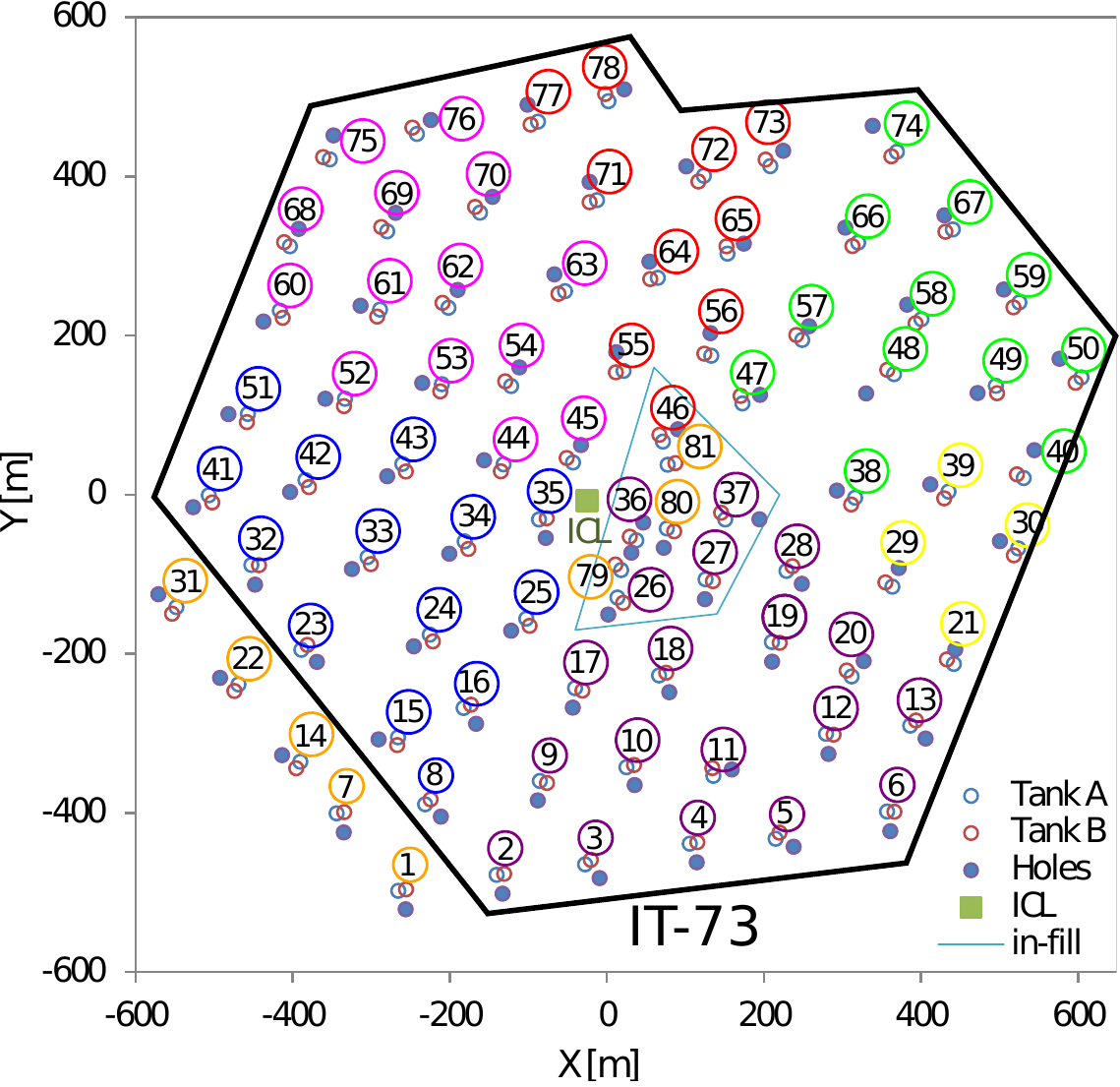}
\end{subfigure}
\hfill
\begin{subfigure}[t]{0.45\textwidth}
\includegraphics[width=\textwidth]{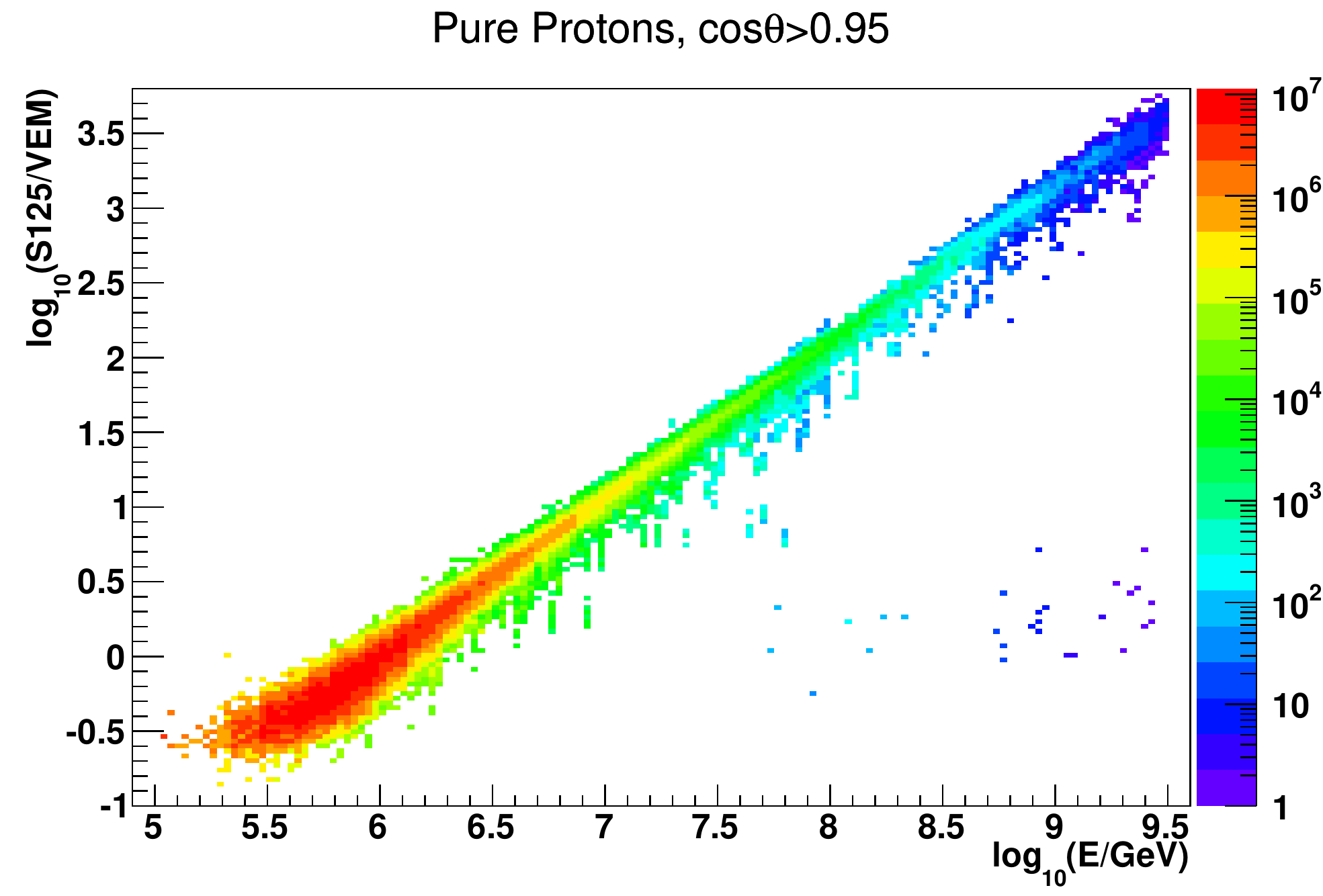}
\end{subfigure}
\hspace*{\fill}
\caption{\footnotesize{\textbf{Left:}  A top view of the IceTop surface array. Colors indicate the construction periods for the strings and tanks. This work will focus on IceTop-73 (IT-73) and IceCube-79 (IC-79), bordered in black \cite{Feusels:thesis}.
\textbf{Right:} Relationship between \s~and primary energy in IceTop for primary protons at high ($\cos(\theta) \ge 0.95$) zenith angles \cite{Andeen_UHECR}.}}
\label{fig:IceTop}
\end{figure}

The \emph{IceCube-InIce array} is the largest neutrino detector in the world: 86 detector strings are instrumented with 60 digital optical modules (DOMs) apiece between 1450~m and 2450~m beneath the surface of the ice sheet, comprising a detector volume of $\sim$1~km$^3$ \cite{IceCube_detector}.  The DOMs are designed to detect the Cherenkov light emitted by charged particles traversing the ice \cite{Abbasi:2010, Abbasi:2009}.  The detector strings are arranged in a triangular grid with $\sim$125~m separation, as shown in Figure \ref{fig:IceTop}~(Left).  

The largest background to the neutrino analyses performed using the IceCube-InIce array are the plentiful high-energy muon bundles created near the first interaction of primary cosmic ray particles with the atmosphere.  Only those muons with sufficient energy at production ($\sim$500~GeV) reach the deep array.  This energy threshold increases with the amount of ice the muons must penetrate and, consequently, with zenith angle.  The number of muons in these bundles is strongly dependent on the composition of the cosmic rays: cosmic ray primaries with more nucleons produce more high-energy muons per shower, higher in the atmosphere, than light primaries of the same energy (mainly due to the superposition principle as discussed in \cite{KAMPERT2012660}, for example).   

Most of the IceCube-InIce detector strings are topped with a surface station comprised of two tanks separated by 10~m \cite{icetop_technical}.  Each tank is an ice-Cherenkov detector viewed by two DOMs apiece--one operating at low gain, the other at high gain--to maximize the dynamic range of the detector.   These surface stations together are called the \emph{IceTop array}.  IceTop is primarily sensitive to the electromagnetic component of incoming cosmic ray air showers, which allows for a nearly composition independent reconstruction of the primary energy of the air showers.  

The IceTop and the IceCube arrays can be operated independently or in coincidence; in this work, two separate analyses are presented (which have also been discussed in \cite{Andeen_UHECR}).  The first is an analysis of the cosmic ray energy spectrum using \textit{IceTop-alone}, and the second is an analysis of the energy spectrum and composition of cosmic rays using \textit{IceTop and IceCube in coincidence}.  Both analyses use the same dataset, Monte Carlo simulations, and IceTop reconstruction parameters.  The coincidence analysis utilizes the additional information from the muon bundles detected by the InIce array.

\section{Data and Simulation}
\label{sec:data_sim}
In these two analyses, three years of IceTop data (2010-2013) are used.  During the first year of this dataset (2010-2011) the full array was not yet completed (this was called IT-73/IC-79, with 73 of 81 stations fully operating at the surface, and 79 of the 86 strings fully operating in the ice \cite{it73_icetopalone_spectrum}); thus the data from the following two years with the complete 81 station / 86 string array is retriggered using only the IT-73 tanks and IC-79 strings for consistency across the three years (as shown in Figure \ref{fig:IceTop}~(Left)).  Monte Carlo simulations of four cosmic ray primary types (proton, helium, oxygen and iron) are also generated to an $E^{-1}$ spectrum using the IT-73/IC-79 detector configuration.  The CORSIKA \cite{CORSIKA_Heck, CORSIKA_manual} air shower generator was used to produce 42000 air showers of each particle type, with FLUKA \cite{FLUKA:2006} as the low-energy hadronic interaction model (below 80 GeV) and SYBILL 2.1 \cite{Ahn:2009} as the high-energy interaction model (above 80 GeV).  A detailed surface detector simulation is implemented in Geant4 \cite{GEANT4:2003, GEANT4:2006}, which models the tank response and the effect of snow on top of the tanks.  These simulations are also used to determine the efficiency of IceTop as a function of primary energy.

\section{IceTop Reconstruction}
\label{sec:it_reco}
The signals from all tanks and all deep-ice detectors are recorded when a basic trigger is satisfied: six tanks in three IceTop stations must register a signal coincident in time \cite{icetop_technical}.  Next, the IceTop data are passed through a maximum-likelihood algorithm to fit both the shape and normalization of the deposited charges to a lateral distribution function (LDF), which also takes into account fluctuations in the arrival times at the detectors.  This reconstruction algorithm results in a number of fitted parameters: the shower core position ($x, y, z$), the shower direction ($\theta, \phi$), and ($S_{125}, \beta$), where $S_{125}$ is the result of the LDF fit to the signal strength measured in vertical equivalent muons (VEM) at a reference distance of 125~m perpendicular to the shower axis, and $\beta$ is related to the slope of the LDF.  $S_{125}$ is directly related the energy of the primary cosmic ray, as shown in Figure \ref{fig:IceTop}~(Right).  For all events, it is important to note that there is a reduction in the electromagnetic signal due to snow, which accumulates at an average of 20~cm per year on top of the IceTop array.   Thus, a snow correction factor is applied during the likelihood calculation, as detailed in \cite{icetop_technical, it73_icetopalone_spectrum}.

\section{IceTop-Alone Energy Spectrum Analysis and Results}
\label{sec:it_alone}
The all-particle energy spectrum using IceTop alone is derived from the measured $S_{125}$ spectrum that results from the snow-corrected reconstruction algorithm discussed above.  The relationship between $S_{125}$ and primary energy for different groups of nuclei is unfolded using the Monte Carlo simulations (as discussed in \cite{it73_icetopalone_spectrum}): slices are made in $S_{125}$ and the mean primary energy of each slice is calculated.  A conversion function is then developed and applied to $S_{125}$ in experimental data.  Since the relationship between $S_{125}$ and primary energy is dependent on the composition of primary cosmic rays (i.e. the relative abundance of different mass groups vs energy), in the IceTop-alone analysis an assumption must be made to derive this relationship.  Although the composition assumption is the source of one of the main systematic uncertainties in extracting the IceTop-alone energy spectrum, the true energy spectrum should be independent of the zenith angle.  Thus, the $S_{125}$ to primary energy unfolding was performed independently for four bins in zenith angle using various composition assumptions.  
The H4a composition model \cite{Gaisser_H4a} provided the most consistent results across the different zenith bins; therefore, the simulated data are weighted using the H4a composition model prior to the $S_{125}$ to primary energy conversion, and the remaining angular dependence is used as a systematic uncertainty (as discussed in \cite{it73_icetopalone_spectrum}). \begin{wrapfigure}[17]{r}{0.5\textwidth}
\centering
\includegraphics[width=0.5\textwidth]{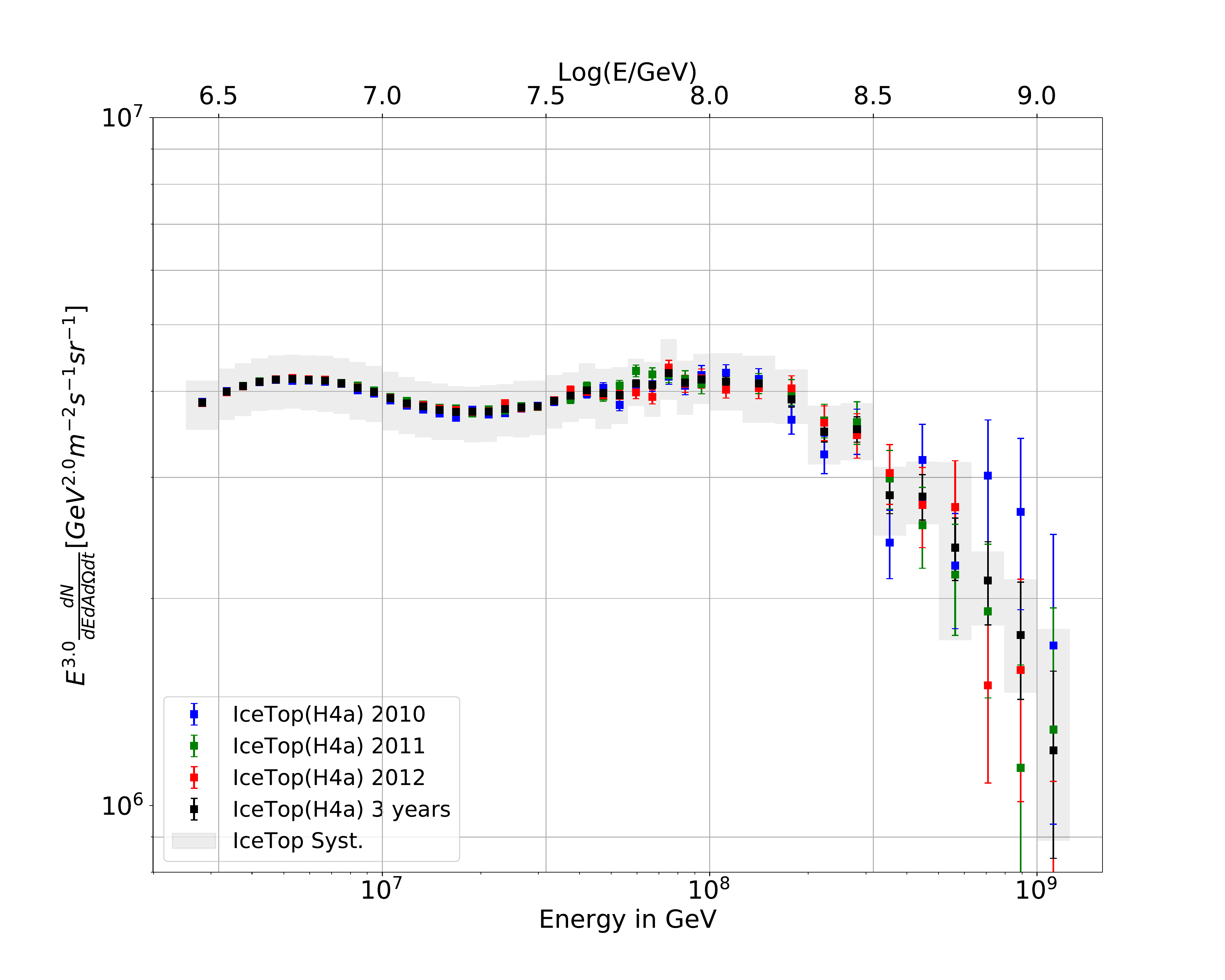}
\caption{\footnotesize{All-particle energy spectrum from the IceTop-alone analysis from each of the three years individually (colors), and all three years together (black) \cite{Andeen_UHECR}.  Systematic uncertainties are shown as the gray band. }} 
\label{f:IT_spectra}
\end{wrapfigure}
After event quality selections (also detailed in \cite{it73_icetopalone_spectrum}), the uncertainty in the direction of events is $\sim$~0.2$^\circ$ at 30~PeV, and the energy resolution for protons at 30~PeV is $\sim$~0.05 in log$_{10}$(E/GeV) \cite{icetop_technical}.  This very good energy resolution is due in part to IceTop's high-altitude location at an atmospheric depth of only $\sim$690~g/cm$^{2}$, which is near the depth of the maximum number of particles for air showers at these energies.  

As shown in Figure \ref{f:IT_spectra}, the IceTop-alone analysis measures the energy spectrum of cosmic rays from $\sim$~300~TeV to $\sim$~2~EeV.  (The low- and high-energy ranges are limited by the spacing of the individual stations and the overall size of the array, respectively.)  The three years of data agree within the statistical and systematic uncertainties.  There is a clear hardening of the spectrum around 2~$\times$~10$^{16}$~eV and a steepening above 2~$\times$~10$^{17}$~eV.

\section{IceCube-InIce Reconstruction}
\label{sec:ITIC_reco}
\begin{figure}[tb]
\centering
\hspace*{\fill}
\begin{subfigure}[t]{0.46\textwidth}
\includegraphics[width=\textwidth]{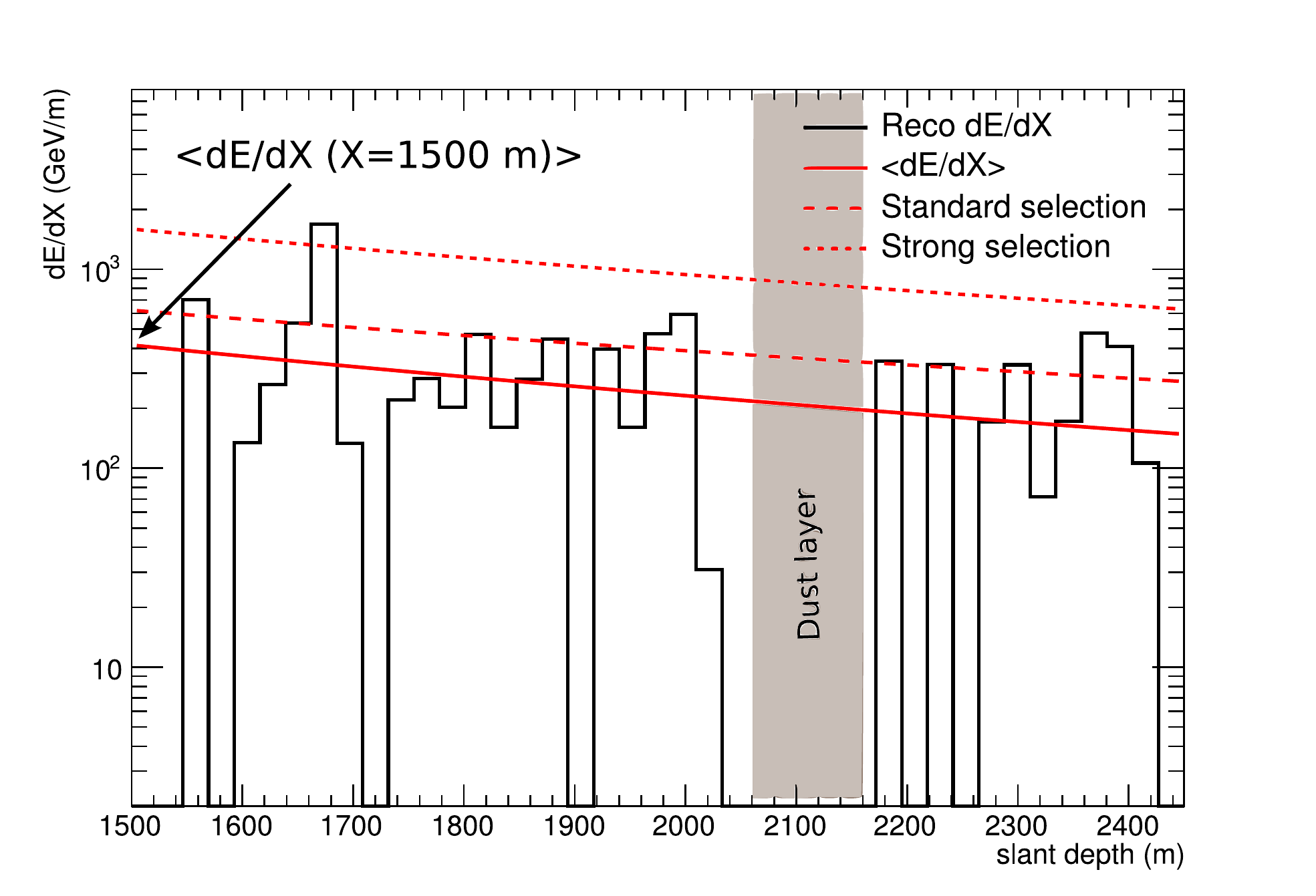}
\end{subfigure}
\hfill
\begin{subfigure}[t]{0.42\textwidth}
\includegraphics[width=\textwidth]{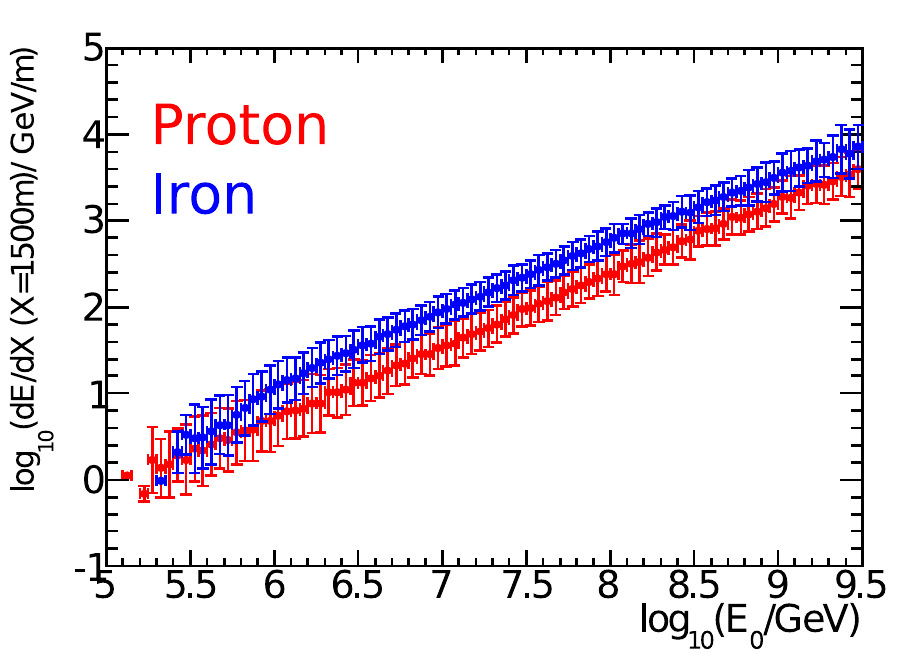}
\end{subfigure}
\hspace*{\fill}
\caption{\footnotesize{\textbf{Left:} Example of the energy loss reconstruction of a large event, where the solid red line demonstrates the average energy loss fit, the dashed red line  represents the standard stochastics selection, and the dotted red line indicates the strong stochastics selection.  The gray band is the approximate location of the dust layer for the slant depth of this particular event. (After \cite{Feusels:thesis, DeRidder:thesis}.)  \textbf{Right:} Composition sensitivity of the energy loss parameter with respect to true primary energy: simulated protons are in red, simulated iron are in blue \cite{Feusels:thesis}.}}
\label{f:composition_reco}
\end{figure}

For events from IceTop that also pass through the IceCube-InIce array, the characteristics of the energy loss of the high-energy muon bundle are also obtained, which provide information about the primary mass of the cosmic rays.  In IceCube-InIce, the pattern of hits for each event is translated into an energy loss profile as a function of slant depth \cite{energy_reco_paper}.  The energy loss profile is then fit to provide three composition-sensitive parameters, as shown in Figure \ref{f:composition_reco}~(Left), \cite{Feusels:thesis}.  First, (dE$_{\mu}$/dX$_{1500}$) is the fitted muon energy loss at X~=~1500~m slant depth, which is a proxy for the total number of muons in the bundle and therefore is sensitive to the mass of the primary cosmic ray.  Two additional parameters are derived from the deviations from the fit, which are caused by stochastic energy losses in the muon bundles.  Since iron-initiated bundles contain more muons for a given energy, they also have more stochastic losses.  On the other hand, since proton-initiated bundles have fewer muons for a given energy, the stochastic losses from proton bundles are more extreme.  A ``strong'' and ``standard'' selection are therefore applied to the fit in order to tease out the composition information from the stochastic energy losses.  The composition-sensitivity of dE$_{\mu}$/dX$_{1500}$ is illustrated in Figure \ref{f:composition_reco}~(Right) from simulations of protons and iron.

\section{IceTop/IceCube Coincident Energy Spectrum and Chemical Composition Analysis and Results}
%UHECR 
\label{sec:ITIC_spectrum_composition}
The same three-year data-set discussed in Section \ref{sec:data_sim} is also used for the IceTop/IceCube coincident analysis of the energy spectrum and the mass composition.  For this analysis, only showers that are successfully reconstructed in IceTop and which pass through the volume of the IceCube-InIce array are preserved.  This selection reduces the amount of data remaining in the final results due to the long lever arm between the two arrays (the remaining data fall within a zenith range of 0~-~30$^{\circ}$); however, the additional muon bundle energy loss information discussed above (Section \ref{sec:ITIC_reco}) is applied to these events.  Therefore, for these events the surface array provides a measurement of the primary energy while the deep IceCube-InIce detector is sensitive to the composition.  Since the energy spectrum and composition are measured in coincidence in this analysis, no composition assumption is necessary to determine the spectrum.

In the IceTop/IceCube coincident analysis, a mass-independent energy spectrum and individual elemental spectra for primary groups are measured using a neural network technique \cite{ic40_coincidence, Feusels:thesis}.  The network is trained on the simulated air showers discussed in Section \ref{sec:data_sim} for all four cosmic ray primary types.  Five input parameters are used for training: the energy proxy $S_{125}$ and the zenith angle from IceTop, and from IceCube-InIce the muon number proxy dE$_{\mu}$/dX$_{1500}$ and two different selections to quantify the high-energy stochastic energy losses along the muon bundle track \cite{Feusels:thesis, DeRidder:thesis}. The network has two outputs: the cosmic ray primary energy and a proxy for the primary mass. 

\begin{figure}[t]%{\textwidth}
\centering
\hspace*{\fill}
\begin{subfigure}[t]{0.54\textwidth}
\includegraphics[width=\textwidth]{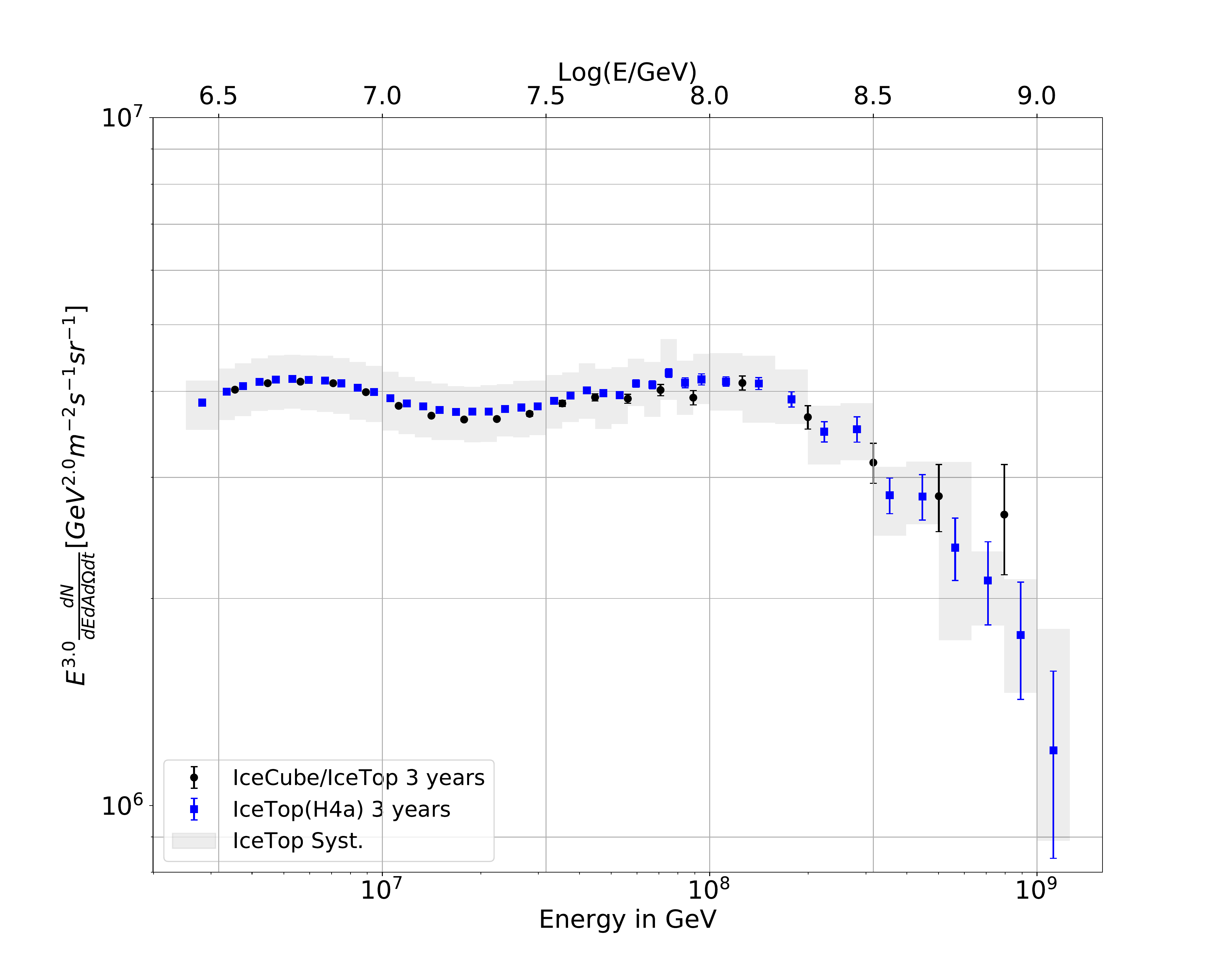}
\end{subfigure}
\hfill
\begin{subfigure}[t]{0.43\textwidth}
\includegraphics[width=\textwidth]{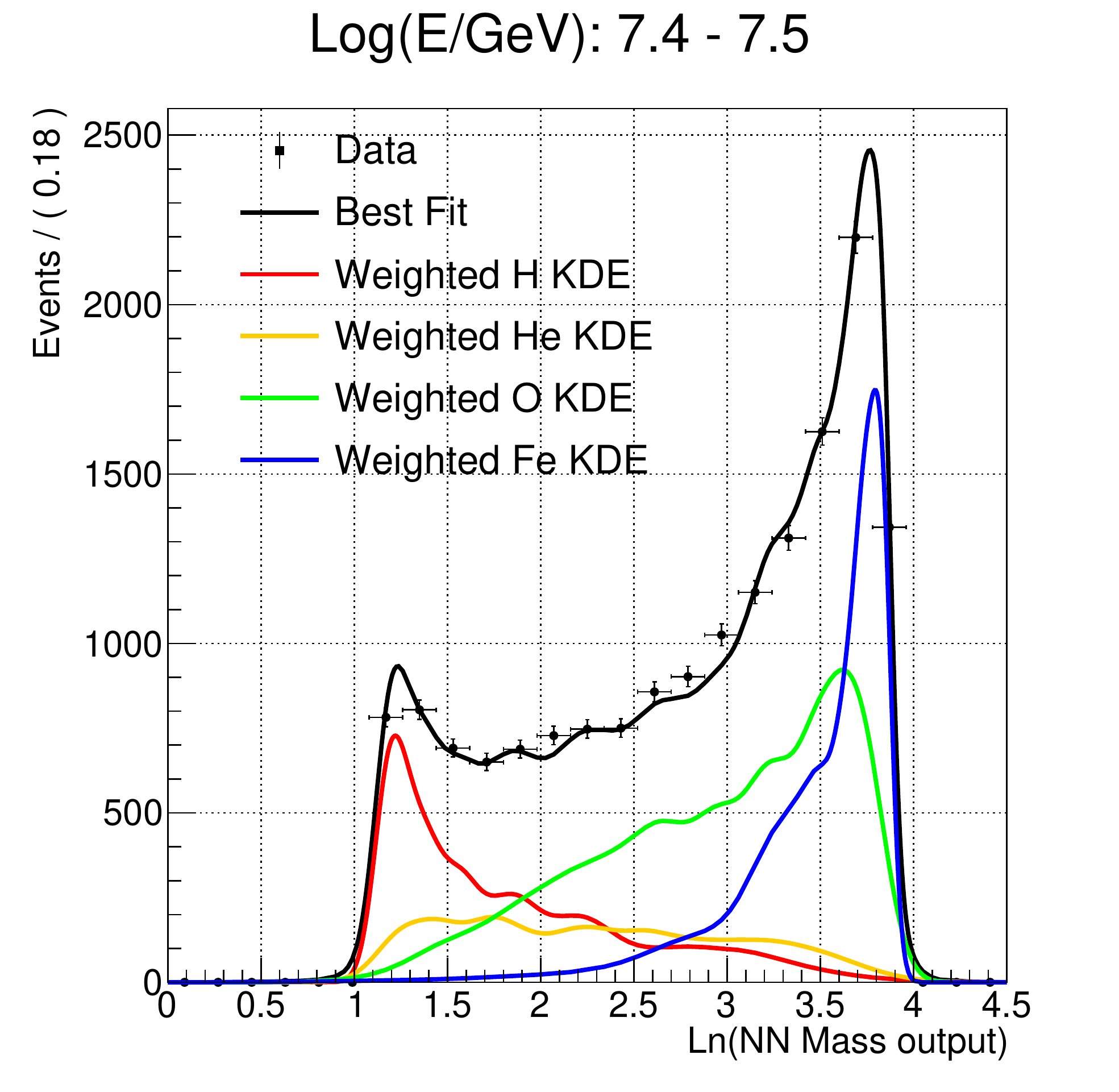}
\end{subfigure}
\hspace*{\fill}
\caption{\footnotesize{\textbf{Left:} A comparison of the combined three-year spectra from the two analyses in these proceedings \cite{Andeen_UHECR}: the IceTop-alone analysis (blue), and the coincident analysis (black). The gray band represents the total systematic uncertainty of the IceTop detector from the IceTop alone analysis. \textbf{Right:} Example simulated mass proxy templates for one slice in energy. The colored lines are the templates derived from the simulated data histograms using the KDE method, scaled to fit the experimental data distribution (black).}}
\label{f:spectra_templates}
\end{figure}

The all-particle cosmic ray energy spectrum is calculated directly from the neural network energy output, taking into account the effective area and livetime of the detector.  The total energy spectrum from the coincident analysis is compared with that from the IceTop-alone analysis in Figure \ref{f:spectra_templates}:~\textbf{Left}.  There is excellent agreement between the two independent analyses.  

The composition as a function of energy is then measured.  The mass proxy distributions are sliced into bins of reconstructed energy.  Within each energy slice, the distribution from each of the four elemental groups simulated (proton, helium, oxygen and iron) is turned into a probability ``template'' using an unbinned kernel density method \cite{KDE_Cramer}.
The experimental data is then compared with the simulated templates to determine the contribution of each of the different mass groups to the data in each slice of reconstructed energy, as shown in Figure \ref{f:spectra_templates}:~\textbf{Right}.  The resulting elemental energy spectra are shown in Figure \ref{f:ElementalSpec}, together with predictions from various recent models.  The measured composition agrees with all predictions within the statistical and systematical detector uncertainties.  The heavy elements maintain a hard spectrum up to higher energies than the lighter elements.

\section{Systematic Effects}
The systematic uncertainties in both the IceTop-alone analysis and the IceTop/IceCube coincident analysis can be grouped into three categories.  Uncertainties in the detectors themselves include the absolute energy scale of IceTop, the snow and atmosphere above IceTop, and the total light yield in the ice (which affects only the coincident analysis).  Analysis choices leading to uncertainties include the choice of binning in the IceTop-alone analysis and the choices made to create the KDE templates in the coincident analysis.  
Finally the choice of hadronic interaction model used for the simulated data presents the greatest hurdle to the composition analysis.

\begin{wrapfigure}[31]{r}{0.5\textwidth}
\centering
\includegraphics[width=.5\textwidth]{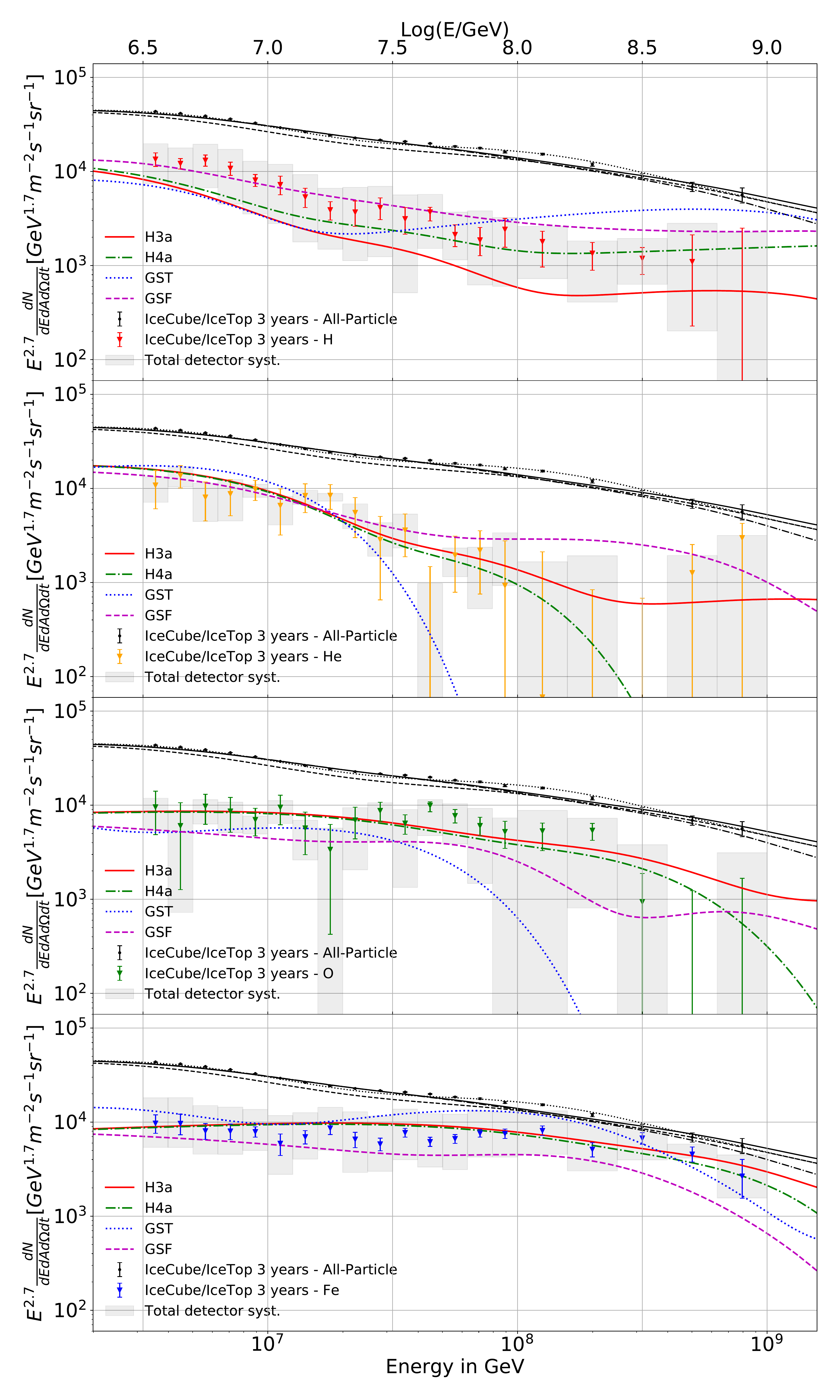}
\caption{\footnotesize{Individual spectra for the four mass groups (protons in red, helium in yellow, oxygen in green, and iron in blue) including total detector systematic compared with various predictions of cosmic ray composition (H3a, H4a, GST, and GSF \cite{Gaisser_H4a, GST_2013, Dembinski:2017zsh}), as shown in \cite{Andeen_UHECR}.}}
\label{f:ElementalSpec}
\centering
\end{wrapfigure}

The detector and analysis effects are estimated, combined, and shown as the grey band in Figures \ref{f:spectra_templates}:~\textbf{Left} and \ref{f:ElementalSpec}.  The largest effect in the IceTop-alone analysis is due to uncertainties in the snow correction ($\pm$ 3\%), while the uncertainty in the light yield in the ice is the largest effect in the coincident analysis (+9.6\% /-12.5\%). 

The uncertainty due to the choice of hadronic interaction model is not calculated using full simulated data samples due to the CPU-time required to generate full samples.  Instead the uncertainty is estimated separately using small samples of three post-LHC hadronic interaction models:EposLHC \cite{EposLHC}, Sibyll2.3 \cite{sibyll2.3}, and QGSJetII-04 \cite{QGSJETII}.  The trend of the all-particle energy spectrum and the composition remain similar, but the choice of hadronic interaction model affects the absolute scale dramatically, particularly in the case of the composition \cite{prd_3yearpaper}.

\section{Discussion}

The two analyses presented here show consistent energy spectrum results, both between each other and between the three years of data individually.  Both analyses show a hardening of the spectrum around 20~PeV in energy, and a softening just above 100~PeV in energy.  At this point, the average composition also changes: at energies up to around 100~PeV, the average mass increases, while above 100~PeV the slope changes and, although the statistical errors become significant here, the average mass could be consistent with either a flat or lightening composition \cite{prd_3yearpaper}.  

In spite of large systematic uncertainties in the absolute scale of the composition results, the trend is consistent: the higher mass elements retain a harder spectrum to higher energies than lighter mass elements.  These spectra are reasonably consistent with the H3a and H4a models \cite{Gaisser_H4a}, and are not inconsistent with the phenomenological GST and GSF models \cite{GST_2013, Dembinski:2017zsh}, although the GST model seems to deviate outside the systematic uncertainty of our results, as shown in Figure \ref{f:ElementalSpec}.  

Both analyses discussed here are forecast to be updated to include more years of experimental data, updated simulations from more intermediate elements, additional composition-sensitive parameters, and results from new internal studies to reduce the detector systematic uncertainties.  These updates are expected to improve the precision of both analyses, and enable the extension of the analyses to higher and lower energies.

\bibliographystyle{ICRC}
\bibliography{references}

\end{document}